\documentclass[amsmath,superscriptaddress]{revtex4-2}
\usepackage{graphicx}
\usepackage{amsmath}
\usepackage{amssymb}
\usepackage{epsfig}
\usepackage{amsthm}
\usepackage{color}
\usepackage{epstopdf}

\def\dd{{\rm d}}
\def\rr{{\bf r}}

\def\abs#1{{{\scriptstyle|}#1{\scriptstyle|}}}

\def\be{\begin{equation}}
\def\ee{\end{equation}}
\def\arr{\begin{array}{rll}}
\def\ea{\end{array}}
\def\bea{\begin{eqnarray}}
\def\eea{\end{eqnarray}}

\begin{document}
\title{ Zernike system revisited: imaginary gauge and  Higgs oscillator }
 
\author{Vahagn Abgaryan}
\email{vahagnab@googlemail.com}
\affiliation{Yerevan Physics Institute, 2 Alikhanian Brothers St., 0036, Yerevan, Armenia}
\affiliation{Joint Institute for Nuclear Research, 6 Joliot-Curie St., Dubna, 141980,  Russia}

\author{Armen Nersessian}
\email{arnerses@yerphi.am}
\affiliation{Yerevan Physics Institute, 2 Alikhanian Brothers St., 0036, Yerevan, Armenia}
\affiliation{Institute of Radiophysics and Electronics, Ashtarak-2, 0203, Armenia}
\affiliation{Joint Institute for Nuclear Research, 6 Joliot-Curie St., Dubna, 141980,  Russia}

\author{Vahagn Yeghikyan}
\email{vahagn.yeghikyan@gmail.com}
\affiliation{Yerevan Physics Institute, 2 Alikhanian Brothers St., 0036, Yerevan, Armenia}

\begin{abstract}
We analyze that recently proposed clasical/quantum mechanical interpretation of Zernike system
and establish its equivalence to the Higgs oscillator on sphere or pseudosphere (Lobachevsky plane).
We show that the non-reality of the classical Zernike Hamiltonian is an insignificant artifact of
imaginary gauge and can be eliminated  with a canonical transformation. The quantum counterpart of this canonical transformation is a similarity transformation mapping the system to the quantum Higgs
oscillator with integration measure depending on $\alpha,\beta$ parameters. When $\alpha=2 \beta$ it results in the
Hermitian Hamiltonian describing a free particle on (pseudo)sphere, while deviation from this point leads to a pseudo-Hermitian system.
\end{abstract}

\maketitle
\section{Introduction}
In 1934, Fritz Zernike proposed a differential equation to classify wavefront aberrations in a circular pupil, where the boundary value can be nonzero \cite{Zernike34}. 
Since its invention, the Zenike system has attracted wide interest  and up to now    it remains a common tool in the study of numerous optical phenomena \cite{Born1999}. It attracts significant interest  in math-physics community  as well (see, e.g.,   \cite{mp,pogosyan1,pogosyan2}).

The differential operator and eigenvalue equation of the Zernike system read
\be 
    \hat Z^{(\alpha,\beta)}\Psi(\rr) := \Big(\boldsymbol{\partial}^2 + 
            \alpha(\rr\cdot\boldsymbol{\partial})^2 
            + \beta\,\rr\cdot\boldsymbol{\partial} \Big)
                    \Psi(\rr) = -E\, \Psi(\rr),\qquad \mathbf{r}:=(x^1,x^2),\quad \boldsymbol{\partial}:=(\partial_1,\partial_2),
                    \label{Zernikeq}
\ee
where $\alpha$ and $\beta$ are real parameters. The Zernike system  is defined on the Hilbert space of square-integrable functions 
 over the unit disk $\mathcal{D} := \{\abs{\rr} \le 1\}$ with the inner product:
\be 
    (f,g)_{\mathcal{D}} := \int_{\mathcal{D}} \dd^2\rr\, f^*(\rr) g(\rr), 
    \label{Z-Hilbert-space}
\ee
where the asterisk denotes complex conjugation, and the functions are required to satisfy the boundary condition $|f(|\mathbf{r}|=1)| = \text{const}$. To ensure the self-adjointness of the operator \eqref{Zernikeq}, Zernike constrained the parameters to $\alpha = -1$ and $\beta = -2$.

Some  decade ago G. Pogosyan, K.B.Wolf et al, revisited   Zernike differntial operator treating it as a quantum mechanical system with a  Hamiltonian   \cite{pogosyan1}:
\be
\widehat{\mathcal{H}} := -\hbar^2 \hat Z^{(\alpha,\beta)} = \widehat{\boldsymbol{\mathfrak{p}}}^2 + \alpha\left(\rr\cdot\widehat{\boldsymbol{\mathfrak{p}}}\right)^2 - \imath {\tilde\beta}\,\rr\cdot\widehat{\boldsymbol{\mathfrak{p}}}, \qquad \widehat{\mathcal{H}}\Psi(\rr) = \hbar^2 E\Psi(\rr), \qquad \widehat{\boldsymbol{\mathfrak{p}}} := -\imath\hbar\boldsymbol{\partial},\qquad {\tilde\beta}:=\hbar\beta
\label{HZ} 
\ee
and interpreted  $\widehat{\boldsymbol{\mathfrak{p}}}$ as the operator of canonical momentum. 

Despite the explicitly non-Hermitian  form of  Hamiltonian \eqref{HZ}   it  has real eigenvalues,  leading the authors to classify it as a $PT$-symmetric Hamiltonian. They also presented its     classical (``de-quantized") counterpart  given by  complex Hamiltonian whose equations of motion admit real solutions \cite{pogosyan2}. Furthermore, they found that this system is superintegrable for arbitrary values of $\alpha$ and $\beta$  and has symmetry algebra of the Higgs oscillator \cite{Higgs}. 
These papers spurred further investigations into this topic, which continue to the present day (see \cite{Zother} and references therein). 

In this note, we demonstrate that the generalized classical  Zernike system  can be transformed,  by the appropriate canonical transformation, to a system with a real Hamiltonian. For $\alpha < 0$, the resulted system is the Higgs oscillator on a two-dimensional sphere of radius $r_0 = 1/\sqrt{|\alpha|}$ and frequency ${\tilde\beta}/2$, while for $\alpha > 0$, it is the Higgs oscillator on the pseudosphere (Lobachevsky plane) of radius $r_0 = 1/\sqrt{\alpha}$ with the same frequency ({\sl Section 2}). 
The quantum counterpart of this canonical transformation is realized through a similarity transformation.
It transforms the quantum Zernike system to a pseudo-Hermitian system with the Hamiltonian of Higgs oscillator on the (pseudo)sphere of radius $r_0 = 1/\sqrt{\alpha}$ and frequency $({\tilde\beta}-2\hbar\alpha)/2$. At the original Zernike point $\beta=2\alpha$ we get the Hermitian Hamiltonian  describing a  free particle on (pseudo)sphere  ({\sl Section 3}). 
We summarize our results and discuss further developments in {\sl Section 4}. 

\section{Classical Zernike System}
The classical (``de-quantized")   Zernike system is given by the canonical Poisson brackets and complex Hamiltonian \cite{pogosyan2}:
\be
\mathcal{H} = \mathbf{p}^2 + \alpha\left(\rr\cdot\mathbf{p}\right)^2 - \imath {\tilde\beta}\,\rr\cdot\mathbf{p}, \qquad \{p_i, x^j\} = \delta_i^j,\quad
\quad \{p_i, p_j\} = \{x^i, x^j\} = 0.
\label{4}\ee

Let us re-write it as  
\be
\mathcal{H} = \boldsymbol{\pi}^2 + \alpha\left(\rr\cdot\boldsymbol{\pi}\right)^2 + \frac{ {\tilde\beta} ^2 \mathbf{r}^2}{4\left(1 + \alpha \rr^2\right)}, \qquad \text{where} \quad \boldsymbol{\pi} \equiv \mathbf{p} - \imath\frac{{\tilde\beta}\mathbf{r}}{2(1 + \alpha \mathbf{r}^2)}.
\label{5}\ee  
In this form, the classical Zernike Hamiltonian may be interpreted as a system coupled with magnetic field defined by the vector potential
\be 
\boldsymbol{\partial} \varphi_{{\tilde\beta}}(\mathbf{r})= \imath\frac{{\tilde\beta}\mathbf{r}}{2(1 + \alpha \mathbf{r}^2)}  , \qquad \varphi_{{\tilde\beta}}(\mathbf{r}) = \imath \frac{{\tilde\beta}}{4\alpha}\log(1 + \alpha\mathbf{r}^2),
\label{vp}
\ee
The imaginary part of the Hamiltonian arises from this vector potential, which is a pure  gauge  and  can be removed via an appropriate canonical transformation. As a result, we get the system with real Hamiltonian
\be
\left(\mathbf{r}, \mathbf{p} - \boldsymbol{\partial} \varphi_{{\tilde\beta}} \right) \to \left(\mathbf{r}, \mathbf{p}\right) : \qquad \mathcal{H} = \mathbf{p}^2 + \alpha(\rr\cdot\mathbf{p})^2 + \frac{{\tilde\beta}^2 \mathbf{r}^2}{4\left(1 + \alpha \mathbf{r}^2\right)}.
\label{canonical}
\ee
 Following the standard procedure of identifying the bilinear form with respect to momentum with a metric (see, e.g., \cite{DeWitt}), one can treat the configuration space as a manifold defined by
\be
ds^2 = \sum_{i,j=1}^2 g_{ij}dx^idx^j=d\mathbf{r} \cdot d\mathbf{r} - \frac{\alpha (\mathbf{r} \cdot d\mathbf{r})^2}{1 + \alpha\mathbf{r}^2},\qquad  {\rm det}\;g_{ij}:=g=\frac{1}{ 1 + \alpha\mathbf{r}^2 }.
\label{metrics}
\ee
For $\alpha < 0$, Eq.~(\ref{metrics}) corresponds to a sphere of radius $r_0 = 1/\sqrt{|\alpha|}$ while  for $\alpha > 0$, it is the metric on the Lobachevsky plane (pseudosphere, the upper sheet of a two-sheet hyperboloid) with the same radius. These metrics  can be embedded  in ambient   three-dimensional Euclidian/Minkowski  spaces:
\be
ds^2 =\sum_{i,j=1}^2 g_{ij}dx^idx^j=d\mathbf{r}\cdot d\mathbf{r} -\text{sgn}(\alpha)  (dx^0)^2, \quad \mathbf{r}^2 -\text{sgn}(\alpha) (x^0)^2 = -\frac{1}{\alpha}.
\ee
The potential appearing in the Hamiltonian \eqref{5} is just   the potential of the (pseudo)spherical Higgs oscillator with frequency $\omega = {\tilde\beta}/2$ \cite{Higgs}:
\be
V_{{\tilde\beta}} = \frac{ {\tilde\beta} ^2 \mathbf{r}^2}{4(1 + \alpha \mathbf{r}^2)}
= \frac{ {\tilde\beta} ^2 r_0^2 \mathbf{r}^2}{4x_0^2},\quad {\rm where}\quad r^2_0=\frac{1}{|\alpha|}
.
\label{V}\ee
Thus, the de-quantized Zernike system is  the Higgs oscillator and its  superintegrability is not surprising.

\section{Quantum Zernike System}
In previous Section we have seen that configuration space of  classical  Zernike Hamiltonian is (pseudo)sphere. It turns out, that the momentum operator $\widehat{\boldsymbol{\mathfrak{p}}}$ defined  in \eqref{HZ} is non-Hermitian in the case of non-constant metrics. To overcome this issue, we define the canonical momentum operator using the following expression \cite{DeWitt}:
\be
\widehat{\mathbf{p}}=-\imath\hbar\left(\boldsymbol{\partial}+\frac{1}{2}\boldsymbol{\partial}\log \sqrt{g}\right)
=\widehat{\boldsymbol{\mathfrak{p}}}+\frac{\imath\hbar\alpha\,\mathbf{r}}{2(1+\alpha\mathbf{r^2})},
\ee
where   $g$  is  defined in \eqref{metrics}.
 
 The   Hamiltonian \eqref{HZ} can be expressed as:
\be 
    \widehat{\mathcal{H}} = 
    \widehat{\boldsymbol{\pi}}^2 + \alpha\left(\widehat{\boldsymbol{\pi}}\cdot\rr\right) \left(\rr\cdot\widehat{\boldsymbol{\pi}}\right)
 + V_{{\tilde\beta}-2\hbar\alpha}(\rr) +\hbar ({\tilde\beta}-2\hbar\alpha),\qquad \widehat{\boldsymbol{\pi}} :=\widehat{\mathbf{p}} -\boldsymbol{\partial}\varphi_{\tilde{\beta}-\hbar\alpha}
 \ee
with $\varphi_{\tilde{\beta}-\hbar\alpha}$  and $V_{{\tilde\beta}-2\hbar\alpha}$   defined by \eqref{vp} and \eqref{V} respectively. 
Similar to  the classical case, the non-Hermitian part of the Hamiltonian arises from the vector potential  which is pure gauge and  
  can be removed at the quantum level as well. Since the gauge is   imaginary function, the quantum counterpart of the canonical transformation \eqref{canonical} is not a unitary transformation but a similarity one, and yields the change of integration measure as well:
\be
{\rm e}^{\frac{\imath}{\hbar} \varphi_{\tilde{\beta}-\hbar\alpha}}\widehat{\boldsymbol{\pi}} 
 {\rm e}^{-\frac{\imath}{\hbar} \varphi_{\tilde{\beta}-\hbar\alpha}}\to \widehat{\mathbf{p}}, \qquad {\rm e}^{\frac{\imath}{\hbar} \varphi_{\tilde{\beta}-\hbar\alpha}}\Psi(\rr) \to \widetilde{\Psi}(\rr),\qquad d^2\mathbf{r}\quad\to\quad {\rm e}^{-\frac{2\imath}{\hbar}  \varphi({r})}d^2\mathbf{r}=\left(1+\alpha  \rr^2\right)^{\frac{\alpha -\beta }{2 \alpha }}d^2\mathbf{r}\,.
 .
\ee
The Hamiltonian then read
\be
\widehat{\mathcal{H}} = 
\widehat{\mathbf{p}}^2 + \alpha\left(\widehat{\mathbf{p}}\cdot\rr\right) \left(\rr\cdot\widehat{\mathbf{p}}\right)  + \frac{ ({\tilde\beta}-2\hbar\alpha)^2 \mathbf{r}^2}{4(1 + \alpha \mathbf{r}^2)}+\hbar ({\tilde\beta}-2\hbar\alpha).
\label{Hf}\ee
%
So,  we arrived at the Hamiltonian of quantum Higgs oscillator   with the frequency $(\tilde{\beta}-2\hbar\alpha)/2$. 
Only at the parameter values $\beta=2\alpha $ (equivalently, $\tilde{\beta}=2\hbar\alpha$) the integration measure 
becomes the expected invariant measure proportional to $\sqrt{g}$  defined in \eqref{metrics}. In this case $\widehat{\mathbf{p}}$ becomes self-conjugated operator, while the potential vanishes. Therefore, we get a  free particle on (pseudo)sphere.
Away from this point  we get the pseudo-Hermitian  system with the Hamiltonian of  Higgs oscillator.

\section{Discussion and outlook}
Let us summarize our results
\begin{itemize}
\item We have shown, that the imaginary part of the classical Zernike Hamiltonian  may be removed by an appropriate canonical transformation induced by a purely imaginary gauge field; the resulting Hamiltonian system is just the Higgs oscillator on a (pseudo)sphere. The role of the parameters is  in defining  the inverse radius of the (pseudo)sphere ($\sqrt{\alpha}$) and the frequency of oscillator  ($\hbar\beta/2$).

\item In the quantum setup, the analogue of the gauge transformation above is a similarity transformation, i.e. ''unitary" transformation with an imaginary phase. This transformation leads the initial Hamiltonian to a visibly  Hermitian form. However, the resulting integration  measure differs from $\sqrt{g}d^{2}\rr$. Thus the Hamiltonian is rendered as pseudo-Hermitian. 
\item  When the Zernike  parameters are $\beta=2\alpha$ the system becomes equivalent to a free particle on a half-(pseudo)sphere with  unaltered volume element and constant boundary condition on the rim, i.e., we get a Hermitian system with a Hamiltonian $\mathcal{H}=\widehat{p}_{i} g^{ij} \widehat{p}_{j}$.

\item The conventional quantization of  Higgs oscillator assumes the replacement of classical kinetic term by the Laplasian on (pseudo)sphere.
 In this terms  the Hamiltonian \eqref{Hf} takes the following form\be
  \widehat{\mathcal{H}} = -\hbar^2\Delta_g +\frac{(\tilde{\beta} -\hbar  \alpha ) (\tilde{\beta} -3 \hbar  \alpha ) r^2}{4 \left(1+\alpha  r^2\right)}+\hbar  (\tilde{\beta} -\hbar  \alpha ),\qquad {\rm where}\quad \Delta_g = \frac{1}{\sqrt{g}} \partial_i  \sqrt{g} g^{ik} \partial_j   = \boldsymbol{\partial}^2 + \alpha (\rr\cdot\boldsymbol{\partial})^2 + \alpha (\rr\cdot\boldsymbol{\partial}).
   \ee
While, we prefered the definition of the kinetic term as $\widehat{p}_{i}g^{ij}\widehat{p}_{j}$.
\end{itemize}
The proposed method can be directly extended to higher-dimensional Zernike systems.

We feel, that beyond the presented results several observations are due regarding the interpretation of Zernike system.
In this work we depart from the ``dequatized" version of the Hamiltonian presented in~\cite{pogosyan2}, obtained
by the set of replacement  rules $\widehat{\mathbf{r}}\leftrightarrow \mathbf{r}$,
 $\widehat{\mathbf{p}}\leftrightarrow \mathbf{p}$ together with $\widehat{\mathbf{r}}\cdot\widehat{\mathbf{p}}$ mapped to $\mathbf{r}\cdot\mathbf{p} $. On the other hand, a more widespread conventions, e.g., Wigner-Weyl approach to the mappings from the observables to functions over the phase space dictate the well-known correspondence $\mathbf{r}\cdot\mathbf{p} \leftrightarrow\frac{\widehat{\mathbf{r}}\cdot\widehat{\mathbf{p}}+\widehat{\mathbf{p}}\cdot\widehat{\mathbf{r}}}{2}$. Adoption of the later  convention leads to the following Hamiltonian
 \be
 \mathcal{H}=\mathbf{p}^2 + \alpha\left(\rr\cdot\mathbf{p}\right)^2 +\imath \hbar\left(2\alpha-\beta\right)\rr\cdot\mathbf{p} + \hbar^{2}(\beta-\alpha). 
 \ee
 This system has two meaningful $\hbar\to 0$ limits. The first assumes, 
 that $\tilde{\beta}=\hbar \beta$ remains finite under the contraction, thus leading to the Hamiltonian \eqref{4}. While, under the second contraction $\alpha$ and $\beta$ are assumed finite, and the Zernike results in a free particle on a (pseudo)sphere. Thus, in  both cases we  deal with maximally superintegrable systems. Respectively, we can relate them with refraction index profiles 
 provided by perfect imaging and cloaking phenomena. The second one   results in the well-known Maxwell fish-eye profile, while the first one corresponds
 to its recently suggested modification \cite{DGN}.

\acknowledgements 
The authors thank Hakob Avetisyan  for attracting our attention to  Zernike  system  and fruitful discussions,   and Tigran Hakobyan for the interest in work. It is sad that we were not able to discuss this topic with George Pogosyan, who passed away last year.  
This work was partially supported by the Armenian State Committee of Higher Education and Science, projects 23/2IRF-1C003 (V.A.) and 21AG-1C062 (A.N., V.Ye.).


\begin{thebibliography}{99}
\bibitem{Zernike34} F.Zernike, Beugungstheorie des Schneidenverfahrens und Seiner Verbesserten Form der Phasenkontrastmethode, \textit{Physica} \textbf{1}, 689--704 (1934).
\bibitem{Born1999} M.Born and E.Wolf, \textit{Principles of Optics: Electromagnetic Theory of Propagation, Interference and Diffraction of Light}, 7th ed. (Cambridge University Press, 1999), p. 986.

E.C.Kintner, On the mathematical properties of the Zernike Polynomials, \textit{Opt. Acta} \textbf{23}, 679--680 (1976).

Kuo Niu and Chao Tian. "Zernike polynomials and their applications." \textit{Journal of Optics} {\bf 24} (2022),123001.
 
E.Goi,  S.Schoenhardt, and Min Gu. "Direct retrieval of Zernike-based pupil functions using integrated diffractive deep neural networks." \textit{Nature Communications } {\bf 13} (2022),  7531.

\bibitem{mp} A.B.Bhatia and E.Wolf, On the circle polynomials of Zernike and related orthogonal sets, \textit{Math. Proc. Cambridge Phil. Soc.} \textbf{50}, 40-48 (1954).

D.R.Myrick, A Generalization of the radial polynomials of F. Zernike, \textit{SIAM J. Appl. Math.} \textbf{14}, 476--489 (1966).

B.H.Shakibaei and R.Paramesran, Recursive formula to compute Zernike radial polynomials, \textit{Opt. Lett.} \textbf{38}, 2487--2489 (2013).

A.Wünsche, Generalized Zernike or disc polynomials, \textit{J. Comp. App. Math.} \textbf{174}, 135--163 (2005).

\bibitem{pogosyan1} G.S.Pogosyan, C.Salto-Alegre, K.B.Wolf, A.Yakhno, ``Quantum superintegrable Zernike system", \textit{J. Math. Phys.} \textbf{58}, 072101 (2017).

G.S.Pogosyan, K.B.Wolf, A.Yakhno, ``New separated polynomial solutions to the Zernike system on the unit disk and interbasis expansion", \textit{J. Opt. Soc. Am. A} \textbf{34}, 1844--1848 (2017).

N.M.Atakishiyev, G.S.Pogosyan, K.B.Wolf, A.Yakhno, ``Interbasis expansions in the Zernike system", \textit{J. Math. Phys.} \textbf{58}, 103505 (2017);``Spherical geometry, Zernike’s separability, and interbasis expansion coefficients", \textit{J. Math. Phys.} \textbf{60}, 101701 (2019).
\bibitem{pogosyan2} G.S.Pogosyan, K.B.Wolf, and A.Yakhno, ``Superintegrable classical Zernike system", \textit{J. Math. Phys.} \textbf{58}, 072901 (2017).
\bibitem{Higgs} P.W.Higgs, ``Dynamical symmetries in a spherical geometry. 1", \textit{J. Phys. A} \textbf{12}, 309--323 (1979).

H.I.Leemon, ``Dynamical Symmetries in a Spherical Geometry. 2,''
\textit{J. Phys. A} \textbf{12}, 489 (1979)
\bibitem{Zother}
E.Celeghini, M.Gadella, M.A. del Olmo, 
``Zernike functions, rigged Hilbert spaces, and potential applications" 
\textit{J. Math. Phys.} {\bf 60}, 083508 (2019)

A. Blasco, I.Gutierrez-Sagredo and F.J.Herranz, ``Higher-order superintegrable momentum-dependent Hamiltonians on curved spaces from the classical Zernike system"
\textit{Nonlinearity}, {\bf 36},  2,  1143 (2023)
 
 
C.Gonera, J.Gonera and P.Kosiński, ``On the generalization of classical Zernike system"
\textit{Nonlinearity}, {\bf  37},  2, 025019 (2024)
 
 
R. Campoamor-Stursberg, F.J.Herranz, D.Latini, I.Marquette, A.Blasco
``Generalized quantum Zernike Hamiltonians: Polynomial Higgs-type algebras and algebraic derivation of the spectrum"
\textit{arXiv:2502.02491 [quant-ph]}

\bibitem{DeWitt}Bryce Seligman DeWitt, ``Point Transformations in Quantum Mechanics", \textit{Phys. Rev.} \textbf{85}, 653 (1952);

D.~McMullan and J.~Paterson,
``Covariant Factor Ordering of Gauge Systems Using Ghost Variables. 1. Constraint Rescaling,''
\textit{J. Math. Phys.} \textbf{30}, 477 (1989);
``Covariant Factor Ordering of Gauge System Using Ghost Variables. 2. States and Observables,''
\textit{J. Math. Phys.} \textbf{30}, 487 (1989).
%
%
\bibitem{DGN}
Z.~Gevorkian, M.~Davtyan and A.~Nersessian,``Extended symmetries in geometrical optics,''
Phys. Rev. A \textbf{101}, 023840 (2020)

M.~Davtyan, Z.~Gevorkian and A.~Nersessian,
``Maxwell fish eye for polarized light,''
Phys. Rev. A \textbf{104}, no.5, 053502 (2021)


\end{thebibliography}
\end{document}